\documentstyle[12pt,tighten,prd,aps]{revtex}
       \def\a{\alpha}
       \def\b{\beta}
       \def\g{\gamma}
       \def\etabar{\bar{\eta}}
       \def\pitil{\tilde{\pi}}
       \def\omegatil{\tilde{\omega}}

       \def\ut#1{\rlap{\lower1ex\hbox{$\sim$}}#1{}}
       \def\lvect#1{\rlap{\raise1.5ex\hbox{$\rightarrow$}}#1{}}
       \def\rvect#1{\rlap{\raise1.5ex\hbox{$\leftarrow$}}#1{}}
       
       \def\half{{\textstyle{1\over2}}}
       \def\IP#1#2{\langle\, #1\, |\, #2\, \rangle}
       \def\ket#1{|\, #1\, \rangle}
       
       \def\real{{\rm I\!R}}
       \def\Tr{\mathop{\rm Tr}}
       \def\D{{\cal D}_a}
       \def\comp{\circ}
       
       \def\G{{\cal G}}
       \def\A{{\cal A}}
       \def\E{{\tilde{\cal E}}}
       \def\F{{\cal F}}
       \def\FG{\F/\G}
       \def\SP{\overline{\FG}}
       \def\Complex{{\bf C}}
       
\input{epsf}

\begin{document}

\preprint{ITP-95- E ,GR-QC/9506029}
\title
{\bf Quantum loop representation for\\
fermions coupled to Einstein - Maxwell field}
\author{Kirill V. Krasnov}
\address
{Bogolyubov Institute for Theoretical Physics, Kiev, Ukraine}
\date{May, 1995}
\maketitle

\bigskip
\centerline{{\it (revised September, 1995)}}

\begin{abstract}

Quantization of the system comprising gravitational, fermionic and
electromagnetic fields is developed in the loop representation. As a result
we obtain a natural unified quantum theory. Gravitational field is treated in
the framework of Ashtekar formalism; fermions are described by two
Grassmann-valued fields. We define a $C^{*}$-algebra of configurational
variables whose generators are associated with oriented loops and curves;
``open'' states -- curves -- are necessary to embrace the fermionic degrees
of freedom. Quantum representation space is constructed as a space of
cylindrical functionals on the spectrum of this $C^{*}$-algebra. Choosing the
basis of ``loop'' states we describe the representation space as the space of
oriented loops and curves; then configurational and momentum loop variables
become in this basis the operators of creation and annihilation of loops and
curves. The important difference of the representation constructed from the
loop representation of pure gravity is that the momentum loop operators act in
our case simply by joining loops in the only compatible with their orientaiton
way, while in the case of pure gravity this action is more complicated.
\end{abstract}

%\narrowtext
\section{Introduction}

Recent developement in nonperturbative quantum gravity has shown that the
loop representation is quite a decent tool for dealing with generally
covariant field theories. This representation allowed one to find a wide class
of solutions of quantum general relativity constraints\cite{RoSm},
\cite{BrPu};there has also been found an interesting interface with the knot
theory \cite{GaPuKnots}.
The main goal of this paper is to construct the loop
representation for the system which includes fermionic and two gauge
fields: gravitational and electromagnetic.

The loop representation in quantum theory is based on using the so-called loop
variables which are the well-known in Yang-Mills theories Wilson loop
functionals. The dynamical variables of Yang-Mills theory (as well as of
general relativity in the framework of Ashtekar variables) are connection
field over the spatial manifold and its conjugate momentum. Wilson loop
functionals form a set of gauge invariant non-local quantities built from
the connection field and the ``loop'' approach is to regard these quantities
as basic variables. This becomes a very powerful means when one considers
a generally covariant field theory. In this case the lack of background
structure does not allow one to construct a renormalized operator
corresponding to a local classical variable -- in other words a
renormalization procedure for constructing such an operator turns out to be
background dependent. On the other hand, loop variables are non-local
quantities; one does not need any background structure to construct a
representation of these variables in terms of operators in a Hilbert space.
So, when there is no background structure available for a quantization
procedure, the following strategy has been proposed\cite{AshtIsham}: it is to
regard loop quantities as basic variables at the classical level and
construct quantum theory representing the corresponding Poisson algebra of
loop variables by a certain operator algebra.

There is also another problem which loop representation seems to be suited for
-- it is the problem of presence of constraints.
Constraints generate symmetry transformations and because of presence of
symmetries not all degrees of freedom of the Lagrange formulation are
physical.
The general strategy for quantizing such a system is to choose the
coordinates on its phase space which have the simplest
properties under the symmetry transformations and regard them as basic
variables. The loop variables are just these quantities.
The symmetry transformations of general relativity in the framework of
Ashtekar variables are gauge transformations and spatial diffeomorphisms. As
we will see, for the system including also fermionic and electromagnetic
fields the symmetry group consists of two similar parts: gauge and
diffeomorphism transformations. It is the advantage of using loop variables
that they are gauge invariant and transform very naturally under the
diffeomorphism group, namely as the geometrical objects with which they are
associated. That is why their usage simplifies considerably the problem
of finding solutions to the gauge and diffeomorphism constraints.

Because loop variables contain {\it all} gauge
invariant information, {\it any} local gauge invariant quantity can be
expressed as a limit of corresponding loop variable. This means that the
Hamiltonian constraint of the theory can be written in terms of loop
variables with a properly chosen limit procedure. This provides us with the
Hamiltonian operator regularization method, because the operator corresponding
to a classical ``loop'' expression becomes a well defined operator in the
loop space. It has been shown by Rovelli and Smolin \cite{RegTechn} that
there exists such a way to take a limit that no divergences will appear
and there arise a well defined operator in the loop space.
So the loop representation
which is based on the usage of loop variables can reduce the problem of
solving the Hamiltonian constraint to a simple combinatorical problem in the
loop space.

As it has been stated above, we develop quantization program for the system
which includes not just pure gravitational field, but also fermionic and
electromagnetic fields. It was noted by Ashtekar {\it et al}
\cite{AsRomTate} that there exists a natural possibility
of unification gravity with other gauge fields in the Hamiltonian
framework; it is to enlarge the gauge group of
pure gravity $SL(2,C)$ to a group which describes
a unified gauge field. The first work along this line \cite{PuUnified}
concerned the loop representation for such a unified theory and its
connection to the knot theory. We continue the
developement of loop representation for the unified theory.
It turnes out that enlargening of the gauge
degrees of freedom, and therefore enlargening of the symmetry
group, leads to some appealing features of the quantum theory in the loop
representation. One of them is that the loop operators act even simpler than
in the case of pure gravity: in the latter case momentum operators act with
a result which includes both a loop and its inverse; this is connected to the
fact that loop variables corresponding to a loop and to its inverse coincide.
In the case of the unified field these two quantities become independent:
the loop variables acquire {\it orientation}. The difference from the case of
pure gravity is that the loop operators never change this orientation when
they act in the ``loop'' space. The Poisson algebra of loop variables is
described solely in terms of breaking, rearranging and rejoining loops and
turns out to be simpler than in the case of pure general relativity.

The example of the loop technique for fermions coupled to gravitational
field was given by Morales-T\'{e}colt and Rovelli \cite{RoMorales}.
Unlike these authors, we consider the full-featured case when two independent
fermionic fields are present. Fermionic fields are described by
two complex Grasmann-valued spinor fields so the ``loop'' variables,
which are mixed ``gauge -- fermionic'' quantities, are even Grassmann algebra
elements. We construct the loop representation in which the action of quantum
analogs of these variables can again be described in a geometric way:
we will see that fermionic operators act in the loop space as
operators of creation and annihilation of curves.

The organization of this paper is as follows. In Sec. \ref{sec:Hf} we remind
briefly the properties of our system in the Hamiltonian formulation and
introduce the unified Einstein-Maxwell gauge field. In this section we also
describe the Hamiltonian formulation for the fermionic system.
In Sec. \ref{sec:LoopVar} we introduce the loop variables and
study the Poisson algebra structure. Sec. \ref{sec:LoopR} is the heart of our
paper: it concerns constructing the loop representation.
We introduce a $C^{*}$-algebra of configurational
variables, find a representation space in which these configurational
variables become operators and, choosing a basis in this space,
define ``loop'' operators.

\section{Hamiltonian formulation}
\label{sec:Hf}

We begin with the action for gravity and matter fields. Fix a four-manifold
$\cal M$, which is topologically a direct product $\real\times\Sigma$ for
some three-manifold $\Sigma$. In the framework of Ashtekar variables the
Lagrangian density for gravity ${\cal L}_E$
is the functional of an anti-Hermitian soldering form $\sigma^{a\;A'}_{\,A}$
and a self-dual
connection ${^4}A_{aA}^{\quad B}$ on $\cal M$
\cite{AsRomTate}
\begin{equation}
{\cal L}_E(\sigma, A) =
       G^2 (\sigma) \sigma^{a\;A'}_{\,A}   \label{GravAction}
       \sigma^{b}_{\;BA'} {^4}F_{ab}^{\ AB},
\end{equation}
where $(\sigma)$ is the determinant of the inverse soldering form and
${^4}F_{ab}$ is the curvature tensor of
${^4}A_{a}$. Self-dual connection field is chosen to be of dimension
$1/${\rm m}, what is convenient because allows one to regard ${^4}\A_{a}$
as a usual Yang-Mills field. Factor $G$ in (\ref{GravAction})
is the fundamental constant; $G$ is set to have a dimension of $1/${\rm m}
so that the action is dimensionless.
Other fundamental constants are set to be $\hbar=c=1$.

Let us note that the action functional (\ref{GravAction}) is
complex; the fields $\sigma^{a}$ and
${^4}A_{a}$ are complex dynamical variables of the complexified general
relativity system.

The connection $\D$ defined by ${^4}A_{aA}^{\quad B}$ and by electromagnetic
vector-potencial ${\bf a}_{a}$ via
${^4}\D\lambda_{A} = \partial_{a}\lambda_{A}+
{^4}A_{aA}^{\quad B}\lambda_{B}+{\bf a}_{a}\lambda_{A}$
acts only on unprimed spinors.
Thus, we shall take the Dirac Lagrangian density for fermionic fields
$\xi^{A}$, $\eta^{A'}$ (Grassmann-valued) to be
\begin{equation}
{\cal L}_D(\xi, \eta, \sigma, A, {\bf a}) =
      {\surd \bar 2}(\sigma)\Bigl [\sigma^{a}\;_{AA'}
      [\bar \xi^{A'}{^4}\D\xi^{A} - ({^4}\D \bar \eta^{A})\eta^{A'}]\;+
      {im\over {\surd \bar 2}}[\bar \eta_{A}\xi^{A}-
                               \bar \xi^{A'}\eta_{A'}]\Bigr ].
\end{equation}

The Lagrangian density for electromagnetic field is
\begin{equation}
{\cal L}_{Em}({\bf a}, \sigma) =
     \half\;(\sigma)g^{ac}g^{bd}\;{^4}{\bf f}_{ab}{^4}{\bf f}_{cd},
\end{equation}
where ${\bf f}_{ab}$ is the curvature tensor of ${\bf a}_a$; metric field
$g^{ab}$ here is defined as the squared soldering form
\[
g^{ab} = \sigma^{a}_{AA'}\sigma^{b\,AA'}.
\]

The total action of the theory is the sum
\[
S = S_{E} + S_{D} + S_{Em}.
\]
In order to develop the canonical quantization program we should
pass on to the Hamiltonian framework, carrying out a space+time
decomposition in the action functional (see \cite{AsRomTate} for details).
Then the action takes the following form
\widetext
   $$S = \int\,dt\int_{\Sigma_t}d^{3}x\Bigl (
     \Tr \tilde{E}^{a} {\cal L}_{t} A_{a}+
     {\cal L}_{t} \xi^{A}\tilde{\pi}_{A}+
     {\cal L}_{t}\bar \eta^{A}\tilde{\omega}_{A}+
     \tilde{{\bf e}}^{a}{\cal L}_{t}{\bf a}_{a}$$
\hbox
 {
 \hskip 0.7in
 \vbox
   {
     \hbox
     {
      $+ \ut{N}\tilde{\tilde{C}}(A, E, \xi, \tilde{\pi}, \bar \eta,
                  \tilde{\omega}, {\bf a}, \tilde{{\bf e}})$
     }
     \bigskip
     \hbox
     {
      $+ N^{a}\tilde{C}_{a}(A, E, \xi, \tilde{\pi}, \bar \eta,
                  \tilde{\omega}, {\bf a}, \tilde{{\bf e}})$
     }
     \bigskip
     \hbox
     {
      $+ ({^4}A\;t)_{B}\,^{A} \tilde{C}_{A}\,^{B}(A, E, \xi, \tilde{\pi},
                  \bar \eta, \tilde{\omega}, {\bf a}, \tilde{{\bf e}})$
     }
     \bigskip
     \hbox
     {
      $+ ({^4}{\bf a}\;t)\;\tilde{c}\;(A, E, \xi, \tilde{\pi}, \bar \eta,
                    \tilde{\omega}, {\bf a}, \tilde{{\bf e}}) \Bigr ).$
     }
   }
  \vbox
   {
    \hbox
      {- Hamiltonian constraint}
    \bigskip
    \smallskip
    \hbox
      {- Diffeomorphism constraint}
    \smallskip
    \hbox
      {- gauge transformations constraint}
    \hbox
      { (spin basis rotations)}
    \smallskip
    \hbox
      {- gauge transformations constraint}
    \hbox
      { (phase rotations)}
    }
 }

%\narrowtext
As it is common for generally covariant field theories, the Hamiltonian will
be the sum of constraints.
The action in this Hamiltonian form provides a canonical phase space
description of the system. This section is aimed to discuss the arising
Poisson structure and to introduce a set of unified coordinates on this
phase space.

\subsection{Einstein-Maxwell unified field}

Let us for the moment restrict our consideration only to the part of the
Hamiltonian descibing the dynamics of the gauge fields.
The last two terms in the Hamiltonian are the generators of local
gauge
transformations on the phase space. These transformations are:
rotations of the complexified spin basis at each
spatial point, which form the group $SL(2,C)$,
and phase rotations, which form the group $U(1)$; the gauge fields lie in the
corresponsing Lie algebras.
Therefore, the full gauge group, which is formed by all internal space
symmetry transformations, is $SL(2,C)\times U(1)$. From
the Hamiltonian, i.e. geometric, point of view it is superfluous to
distinguish the two gauge fields --
the dynamical variables of the theory should be a connection
on some bundle over the spatial manifold (which takes values in the Lie algebra
of the gauge group) and its conjugate momentum. Thus, we should regard the two
independent connection fields of initial Lagrange formulation as the
two parts of one connection field -- the unified
Einstein-Maxwell field.

So we are to choose the new ``coordinates'' on the phase space of the
system which will correspond to the unified gauge field.
The expression for the new gauge variables is straightforward.
Let us choose the new connection field to be
\begin{equation}
{}^4{\cal A}_{aA}^{\quad B} :=
        {}^4A_{aA}^{\quad B} + {}^4{\bf a}_{a}\,\delta_{A}^{\,B}.
                          \label{unified}
\end{equation}
Then the initial Einstein and Maxwell connection fields can be expressed
through ${}^4{\cal A}$ as follows
\begin{eqnarray}
{}^4A_{A}^{\; B} =
         {}^4{\cal A}_{A}^{\; B} - \half(\Tr {}^4{\cal A})\delta_{A}^{\;B},\\
{}^4{\bf a} = \half(\Tr {}^4{\cal A})\delta_{A}^{\; B}.
                       \eqnum{\number\theequation a}
\end{eqnarray}
Having introduced the unified connection field ${\cal A}$ we can
define the corresponding momentum field ${\cal E}$. We shall take it
in the form
\begin{equation}
\tilde{\cal E}^{a\;B}_{\,A} :=
  {\tilde E}^{a\;B}_{\,A} + \half {\bf \tilde{e}}^{a} \delta_{A}^{\,B},
  \label{momentum}
\end{equation}
so that it is the canonically conjugate to ${\cal A}$:
\begin{equation}
\bigl \{ \tilde{\cal E}^{a}_{\,CD}(x), {\cal A}_{b}^{\,AB}(y) \bigr \} =
  -\, \delta^{3}(x-y) \delta_{b}^{\,a} \delta_{D}^{\,A} \delta_{C}^{\,B}.
  \label{brackets}
\end{equation}
Here ${\cal A}$ is the pullback of ${}^4{\cal A}$ to the tree-manifold
$\Sigma$.
The factor $\half$ in (\ref{momentum}) is important;
it provides the correct
(canonical) commutational relations between the connection field and its
momentum (\ref{brackets}).
The gravitational and electromagnetic momentum fields can also be expressed
through the unified field
\begin{eqnarray}
\tilde{E}^{a\;B}_{\,A} = \tilde{\cal E}^{a\;B}_{\,A}\,
     -\,\half \Tr(\tilde{\cal E}^{a})\delta_{A}^{\,B} \\
\tilde{\bf e}^{a} = \Tr(\tilde{\cal E}^{a})
                   \eqnum{\number\theequation a}
\end{eqnarray}

Having these relations it is just an exercise to rewrite the constraints
in terms of the unified fields. The last two terms in the Hamiltonian
are the Gauss law constraints for the gravitational and electromagnetic
fields
\begin{eqnarray}
  ({}^{4}A\dot t)_{A}^{\;B}
                   \; D_{a}\,{\tilde E}^{a\;A}_{\,B}
 +\;({}^{4}{\bf a}\dot t)
                    \partial_{a}{\bf \tilde{e}}^{a}, \nonumber
\end{eqnarray}
where
  $$D_{a} = \partial_{a} + A_{a}.$$
We can express it in terms of the new Lagrange multiplier
$({}^4{\cal A}\dot t)$,
so the Gauss law constraint for the unified field takes the form
\begin{equation}
{\delta (S_{E}+S_{Em})\over
                  \delta ({}^4{\cal A}_{A}^{\,B}\dot t)}
      = \D\tilde{\cal E}^{a\;A}_{\,B} = 0,
\end{equation}
where we have introduced
\begin{equation}
\D := \partial_{a} + {\cal A}_{a}.  \label{D}
\end{equation}

The part of the diffeomorphism constraint
\begin{eqnarray}
{\delta(S_{E}+S_{Em})\over \delta N^{a}} =
     -\;\Tr(\tilde{E}^{b}F_{ab}) - \;{\bf \tilde{e}}^{b}{\bf f}_{ab}
\end{eqnarray}
expressed through the unified variables takes the form
\begin{equation}
{\delta(S_{E}+S_{Em})\over \delta N^{a}} =
  -\;\Tr(\tilde{\cal E}^{b}{\cal F}_{ab}).
\end{equation}
Again, the factor $\half$ from (\ref{momentum}) was necessary to cansel the
factor $2$ which appeared from the trace operation. Here we introduced
the curvature field ${\cal F}_{ab}$ of the connection ${\cal A}_{a}$
\[
  {\cal F}_{ab} := 2\,{\cal D}_{[a}{\cal A}_{b]}
 = 2\,\partial_{[a}A_{b]} + 2\partial_{[a}{\bf a}_{b]} + [A_{a}, A_{b}]
 = F_{ab} + {\bf f}_{ab}.
\]

\subsection{Fermionic part}

For the Dirac action functional the space-time decomposition leads to the
following expression (see \cite{AsRomTate} for details)
\begin{equation}
S_{D} = \int dt\int_{\Sigma_{t}}d^{3}x\;\bigl \{
               -i(\sigma)[(\xi\dagger)_{A}{\cal L}_{t}\xi^{A}\,+
               \,(\etabar\dagger)_{A}{\cal L}_{t}\etabar^{A}]\,+
          \,{\cal H}(\xi, \xi\dagger, \etabar, \etabar\dagger) \bigr \},
          \label{SDinit}
\end{equation}
where ${\cal H}$ means the Hamiltonian functional. The $\dagger$-operation
here descends from the complex conjugation on the Grassmann algebra of
$SL(2,C)$ spinors and satisfies the following properties (a)
$(a\a_{A}+b\b_{A})^{\dagger}=a^{*}\a^{\dagger}_{A}+b^{*}\b^{\dagger}_{A}$;
(b) $(\a^{\dagger}_{A})^{\dagger}=-\a_{A}$; (c)
$(\a^{A})^{\dagger}\a_{A}\geq0$; (d)
$(\epsilon_{\,AB})^{\dagger}=\epsilon_{\,AB}$; (e)
$(\a_{A}\b_{B})^{\dagger}=\a^{\dagger}_{A}\b^{\dagger}_{B}$, for all Grassmann
fields $\a_{A}$ and $\b_{B}$ and complex functions $a, b$.
Being Grassmann-valued,
the fermionic fields anti-commute, so having rearranged them
in (\ref{SDinit}) we got the different from \cite{AsRomTate} sign in
square brackets. We define the momentum
fields by the left variational derivatives
\begin{eqnarray}
\pitil_{A}:={\vec \delta\;S\over\delta\;{\cal L}_{t}\xi^{A}} =
  i(\sigma)(\xi\dagger)_{A}                              \label{momdefine},\\
\omegatil_{A}:={\vec \delta\;S\over\delta\;{\cal L}_{t}\etabar^{A}} =
  i(\sigma)(\etabar\dagger)_{A}.                         \nonumber
\end{eqnarray}

Then the action takes the form
\begin{equation}
S_{D} = \int dt\int_{\Sigma_{t}}d^{3}x\;\bigl [
{\cal L}_{t}\xi^{A}\pitil_{A}\;+\;{\cal L}_{t}\etabar^{A}\omegatil_{A}\;+
\;{\cal H}(\xi, \pitil, \etabar, \omegatil) \bigr ].    \label{DiracAction}
\end{equation}
The momentum fields have appeared at the right side to the configurational
fields because of the usage of the
left derivatives in the momentum field definition. The full Hamiltonian
density for the spinor fields is
\FL
\[
{\cal H}(\xi, \pitil, \etabar, \omegatil) =
\]
\begin{eqnarray}
\ut{N}\;\Bigl [G^{-2}\,{\tilde E}^{a\;B}_{\,A}
[\D\xi^{A}\pitil_{B}\,+\,\D\etabar^{A}\omegatil_{B}]\,+
\,im[(\sigma)^{2}\etabar_A\xi^{A}\,+\,\pitil^{A}\omegatil_{A}] \Bigr ]
                                                           \nonumber \\
                          \smallskip
-\,({}^{4}{\cal A}\;t)_{B}^{\;A}[\xi^{B}\pitil_{A}\,+
                         \,\etabar^{B}\omegatil_{A}]
                                                            \nonumber \\
                          \smallskip
-\,N^{a}[\D\xi^{A}\pitil_{A}\,+\,\D\etabar^{A}\omegatil_{A}],\label{DiracHam}
\end{eqnarray}
where we used the ``unified'' Lagrange multiplier $({}^4{\cal A}\;t)$
(see (\ref{unified})).

The equations of motion are now straightforward from the variational
principle. Using the left variation which turns into zero at the initial
and final time points one finds the dynamics
\[
{\cal L}_{t}\xi^{A} = -\;{\delta\,H\over \delta\,\pitil_{A}}
\quad;\quad
{\cal L}_{t}\etabar^{A} = -\;{\delta\,H\over \delta\,\omegatil_{A}}
\]
\[
{\cal L}_{t}\pitil_{A} = -\;{\delta\,H\over \delta\,\xi^{A}}
\quad;\quad
{\cal L}_{t}\omegatil_{A} = -\;{\delta\,H\over \delta\,\etabar^{A}}.
\]
So the evolution of any functional of the dynamical variables is given by
\[
{\cal L}_{t}f(\xi, \pitil, \etabar, \omegatil) =
 \{\,H\,,\,f\,\},
\]
where the Poisson structure on the phase space is defined via
\begin{equation}
\{\,f\,,\,g\,\} = -\;\int d^{3}x \Bigl [
\,{\delta\,f\over \delta\,\xi^{A}}\;{\delta\,g\over \delta\,\pitil_{A}}\,
+\,{\delta\,f\over \delta\,\pitil_{A}}\;{\delta\,g\over \delta\,\xi^{A}}\,
+\,{\delta\,f\over \delta\,\etabar^{A}}\;
                                  {\delta\,g\over \delta\,\omegatil_{A}}\,
+\,{\delta\,f\over \delta\,\omegatil_{A}}\;
                                  {\delta\,g\over \delta\,\etabar^{A}}\,
                                  \Bigr ].
\end{equation}
All functional derivatives in this formula are left. Then one can obtain
the Poisson brackets between the canonical variables
\begin{equation}
 \{\,\xi^{A}(x),\,\xi^{B}(y)\} = 0\quad;\quad
 \{\,\pitil_{A}(x),\,\pitil_{B}(y)\} = 0;         \nonumber
\end{equation}
\begin{equation}
 \{ \,\pitil_{B}(y),\,\xi^{A}(x) \} = -\,\delta_{B}^{\;A}{\tilde\delta}(x-y)
                                  \label{FermBrackets}
\end{equation}
and analogously for $\etabar, \omegatil$ fields.

\subsection{The Hamiltonian constraint}

The Hamiltonian constraint of the theory consists of the three parts
\begin{eqnarray}
{\delta S_{E}\over \delta \ut{N}} =
\half {1\over G^{2}} \Tr(\tilde{E}^{a}\tilde{E}^{b}F_{ab}) \label{HC}\\
{\delta S_{Em}\over \delta \ut{N}} =
{1\over 32}\,{1\over (G^{2})^{4}} (\sigma)^{-2}
\Tr(\tilde{E}^{a}\tilde{E}^{c})\,\Tr(\tilde{E}^{b}\tilde{E}^{d})
\,\bigl ( {\bf e}_{ab}{\bf e}_{cd}\,
                  \,+ {\bf b}_{ab}{\bf b}_{cd} \bigr )
                                          \eqnum{\number\theequation a}\\
{\delta S_{D}\over \delta \ut{N}} =
{G}^{-2}{\tilde{E}^{a\;B}_{\,A}}
   \bigl (\D\xi^{A}\tilde{\pi}_{B}\,+\,\D\etabar^{A}\tilde{\omega}_{B}\bigr)\,
    +\,im\,\bigl ((\sigma)^{2}\etabar_{A}\xi^{A}\,
               +\,\tilde{\pi}^{A}\omegatil_{A}\bigr ),
                                           \eqnum{\number\theequation b}
\end{eqnarray}
where ${\bf b}_{ab} = 2\;{\bf f}_{ab}$.
Having introduced the unified connection field and the corresponding
conjugate momentum,
we shall express the Hamiltonian constraint in terms of these fields.
This gives for the Einstein part of the Hamiltonian
\begin{equation}
{\delta S_{E}\over \delta \ut{N}} =
\half {1\over G^{2}} \;\ut{\eta}_{\,abc}
\Tr(\tilde{\cal E}^{a}\tilde{\cal E}^{b}\tilde{\cal B}^{c}). \label{Hamilt}
\end{equation}
Here we introduced the magnetic field $\tilde{\cal B}^{a}$ as the dual of
the curvature of the unified field
$${\cal F}_{ab} = \ut{\eta}_{\,abc}\tilde{\cal B}^{c},$$
so that it has the dimension and the weight of $\tilde{\cal E}^{a}$.
The tensor
$\ut{\eta}_{\,abc}$ is the totally antisymmetric tensor of weight $-1$.

The other two parts of the Hamiltonian become
\begin{eqnarray}
{\delta S_{Em}\over \delta \ut{N}} =
{1\over 32}\,{1\over (G^{2})^{4}} (\sigma)^{-2}
\ut{\eta}_{\,abe}
\ut{\eta}_{\,cdf}\, \Tr(\tilde{\cal E}^{a}\tilde{\cal E}^{c})
\Bigl (\Tr(\tilde{\cal E}^{b}\tilde{\cal E}^{d})
      \Tr(\tilde{\cal B}^{e}) \Tr(\tilde{\cal B}^{f}) \nonumber \\
   -\,\Tr(\tilde{\cal E}^{b}\tilde{\cal E}^{d})
      \Tr(\tilde{\cal E}^{e}) \Tr(\tilde{\cal E}^{f}) \,
   -\,\Tr(\tilde{\cal E}^{b}) \Tr(\tilde{\cal E}^{d})
      \Tr(\tilde{\cal B}^{e}) \Tr(\tilde{\cal B}^{f}) \Bigr ),
                      \eqnum{\number\theequation a}
\end{eqnarray}
\begin{eqnarray}
{\delta S_{D}\over \delta \ut{N}} = {G}^{-2}
\bigl (\tilde{\cal E}^{a\;B}_{\,A}\,
-\,\half \Tr(\tilde{\cal E}^{a})\delta_{A}^{B}\bigr )
\bigl (\D\xi^{A}\tilde{\pi}_{B}\,+\,\D\etabar^{A}\tilde{\omega}_{B}\bigr )\,
    +\,im\,\bigl ((\sigma)^{2}\etabar_{A}\xi^{A}\,
       +\,\tilde{\pi}^{A}\tilde{\omega}_{A}\bigr ).
       \eqnum{\number\theequation b}
\end{eqnarray}

Let us also give here the complete (including fermionic degrees of freedom)
expression for the Gauss law and diffeomorphism
constraints in terms of the Einsein-Maxwell field
\begin{eqnarray}
{\delta S\over \delta N^{a}} =
-\;\Tr(\tilde{\cal E}^{b}{\cal F}_{ab})\;
-\;(\D\xi^{A}\pitil_{A}\,+\,\D\etabar^{A}\omegatil_{A}) \\
{\delta S\over \delta ({}^4{\cal A}\;t)_{A}^{\;B}} =
-\;(\xi^{A}\pitil_{B}\,+\,\etabar^{A}\omegatil_{B})\;+
\;\D\tilde{\cal E}^{a\;A}_{\;B}.
\end{eqnarray}

As we have seen, the Hamiltonian constraint contains
the determinant of the inverse soldering form
\begin{equation}
(\sigma)^{2} =
-\;{1\over 3\surd{2}}\;\ut{\eta}\,_{abc}
\Tr(\tilde\sigma^{a}\tilde\sigma^{b}\tilde\sigma^{c}), \label{determ}
\end{equation}
so we need its expression
through the unified variables. It is given by
\begin{equation}
(\sigma)^{2} = {{\rm i}\over 12\,G^{6}}\,\ut{\eta}_{\,abc}
\Tr\bigl (\tilde{\cal E}^{a}\,-\,\half \Tr(\tilde{\cal E}^{a})\bigr )
   \bigl (\tilde{\cal E}^{b}\,-\,\half \Tr(\tilde{\cal E}^{b})\bigr )
   \bigl (\tilde{\cal E}^{c}\,-\,\half \Tr(\tilde{\cal E}^{c})\bigr ).
   \label{determfinal}
\end{equation}

This accomplishes the aim of this Section, which was to obtain all the
constraints of the Hamiltonian framework expressed in terms of the unified
gauge and the fermionic fields. We will conclude by pointing out that
in the form (\ref{Hamilt}) the Hamiltonian is not polynomial in
$\tilde{\cal E}^{a}$ variables because of the presence of the factor
$(\sigma)^{-2}$ in the electromagnetic part;
this might cause problems in constructing the
corresponding quantum operator. Possible solution of the
problem was proposed by Ashtekar {\it et al} \cite{AsRomTate}.
Multiplying the Hamiltonian constraint by $(\sigma)^{2}$ one may
restore its polynomial character; the Hamiltonian constraint
becomes a density of weight four (therefore the corresponding
Lagrange multiplier - lapse function - becomes
a density of weight minus three). Another possible way to tackle this
problem is discussed in \cite{VolumeOp}.

\section{Algebra of Loop Variables}
\label{sec:LoopVar}

In this Section we construct
gauge invariant non-local functionals of the dynamical variables.
These functionals are associated with loops,
curves and ribbons so they will play a role of ``loop'' gauge-invariant
coordinates on the system's phase space.
We will discuss the algebra of these ``loop'' variables with respect to the
Poisson brackets. A special attention is paid to a graphical representation
of the algebra obtained.

\subsection{Configurational loop variables}

The set of variables which we call configurational loop variables will
play a crucial role in the quantization procedure.
Let us denote the space of the unified connection fields (the space of
connections on a certain $SL(2,C)\times U(1)$ bundle over $\Sigma$) by
$\F$ and consider a
Wilson loop functional on $\F$
\begin{equation}
(\g) \equiv T_{\g}[{\cal A}] := \Tr {\cal P}\,\exp{\oint_{\g}{\cal A}},
\end{equation}
$$\g : [0,1] \to \Sigma.$$
The expression under the trace operation here is the parallel transport
(with the connection ${\cal A}$) matrix $U$
\[
U[\g]_{A}^{\;B} =
    {\cal P} \exp{\int_{\g}d\,{\tau}\,{\dot\g}^{a}\,{\cal A}_{aA}^{\quad B}}
\]
taken for a closed loop, so
\[
(\g) = \Tr\,U[\g].
\]

The main difference from the case of pure gravity is that
\begin{equation}
(\g^{-1})[\A]\not=(\g)[\A]   \label{NotEquiv}
\end{equation}
because of the presence of the
additional electromagnetic part in the connection field. Since loop
quantities form a set of (complex) coordinates\footnote{
in the sense that for a pair of gauge
not equivalent fields $\A_{1}$ and $\A_{2}$ there exist such a pair of loops
$\g_{1}$ and $\g_{2}$ that $(\g_{1})[\A_{1}]\not=(\g_{2})[\A_{2}]$}
on the configurational
space $\F/\G$ (we denoted by $\F/\G$ the quotient space of $\F$ with respect
to the action of gauge transformations)
(\ref{NotEquiv}) means that the loop quantities $(\g)$ and $(\g^{-1})$
are independent coordinate variables. This explains why, unlike the case of
pure general relativity, loop variables are associated with {\it oriented}
loops. We will exploit a simple graphical representation of these quantities
\begin{figure}
\moveright 2.5in \vbox{\epsffile{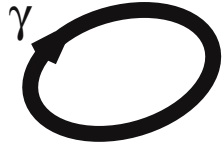}}
\caption{Configurational loop variables for the unified field are
associated with oriented loops.}
\end{figure}

Loop variables form an over-complete set of coordinate quantities in the
sense that they satisfy the following identities \cite{PuUnified}
\begin{enumerate}
\item They are invariant under reparametrizations of loops. If $\g'$ is
a reparametrized loop
$\g'(s) = \g\,(\,f(s)\,)$
then
\[
(\g') = (\g).
\]
\item The Mandelstam identity. For any three loops $\a, \b,$ and $\g$
intersecting at a point one has
\[
(\a)(\b)(\g) = (\a\comp\b)(\g) + (\a)(\b\comp\g) + (\a\comp\g)(\b) -
   (\a\comp\b\comp\g) - (\a\comp\g\comp\b).
\]
This, on the first sight cumbersome relation has replaced the simpler
Mandelstam identity for the case of pure gravity \cite{RoSm}
\[
(\a)(\b) = (\a\comp\b) + (\a\comp\b^{-1})
\]
owing to the independence of $(\g)$ and $(\g^{-1})$ variables in our case.
\item Loop variables are invariant under retracing operation
\[
(\g \comp \eta \comp \eta^{-1}) = (\g),
\]
where $\comp$ means the composition of loops which intersect at a point,
$\eta^{-1}$ is the inverse of a curve $\eta$.
\end{enumerate}

As configurational loop variables involving  fermionic degrees of
freedom we will take certain  even Grassmann algebra elements.
The infinite-dimensional Grassmann algebra is generated by the anticommuting
complex objects --  our dynamical field
variables $\xi(x), \etabar(x), \pitil(x), \omegatil(x)$. A basis of the
Grassmann algebra is formed by the powers of these generators.
Let us consider the following gauge-invariant even elements associated
with open curves
\begin{equation}
(\xi|\g|\bar \eta):= \Tr \bigl \{\xi\,U[\g]\,\bar \eta \bigr \} =
   \xi^{A}\,U[\g]_{A}^{\;B}\,\bar\eta_{B},
\end{equation}
which we will regard as fermionic configurational variables%
\footnote
{This is the point where our approach differs from that of
Morales-T${\acute e}$colt and Rovelli
\cite{RoMorales}. As the quantities involving fermionic fields they
considered $(\psi|\g|\psi)$ (in our notations). It is one of the reasons for
which we introduce the loop variables that the quantum Hamiltonian operator
can be defined as a certain loop limit of an operator constructed from the
basic loop operators. However,
the quantities they consider as basic ones turn into zero when the
corresponding curve srinks to a point. In this sense the loop quantities
quadratic in a Grassmann field can not serve as basic variables.}

We propose a convinient notation in which any ``loop'' quantity is denoted
by a Greek letter in parenthesis. Since ends of a curve correspond
to the fermionic degrees of freedom, it is convinient to include
symbols of fermionic fields in paranthesis on both sides
of a loop symbol to get a symbol which describes
the mixed quantity. Thus, $\g$ in the above expression is the open curve
with ends marked by
$\xi, \bar\eta$; we will always put $\xi$ at the
{\it final} point of a curve and $\eta$ will mark the {\it initial}
(recall that any curve (loop) has an orientation)
\begin{figure}
\moveright 2.5in \vbox{\epsffile{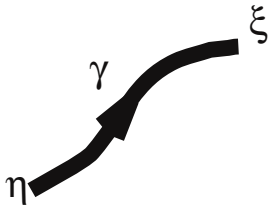}}
\caption{Fermionic variables are associated with open curves.}
\end{figure}

The quantities introduced satisfy the relations analogous to those for
closed loops
\begin{enumerate}
\item Reparametrization invariance
\[
(\xi|\g'|\bar\eta) = (\xi|\g|\bar\eta).
\]
\item The Mandelstam identity.
Consider a curve $\a$ and a point $s$ on it.
This point divides $\a$ into two parts for which we use the special
notation
\[
\a = \a_{/s}\comp\a_{s}.
\]
Thus, $\a_{s}$ is the part of the $\a$ from the initial point to the
point $s$ and $\a_{/s}$ is the remaining part.
Then for any three curves $\a, \b,$ and $\g$
intersecting at a point the following identity holds
\widetext
\FL \[
(\xi|\a|\bar\eta)(\xi|\b|\bar\eta)(\xi|\g|\bar\eta) = \]
\begin{eqnarray}
= (\xi|\a_{/i}\comp\b_{i}|\bar\eta)(\xi|\b_{/i}\comp\a_{i}|\bar\eta)
                                               (\xi|\g|\bar\eta)  \,+\,
(\xi|\a_{/i}\comp\g_{i}|\bar\eta)(\xi|\g_{/i}\comp\a_{i}|\bar\eta)
                                               (\xi|\b|\bar\eta)\\  +\,
(\xi|\b_{/i}\comp\g_{i}|\bar\eta)(\xi|\g_{/i}\comp\b_{i}|\bar\eta)
                                               (\xi|\a|\bar\eta)  \,- \,
(\xi|\g_{/i}\comp\a_{i}|\bar\eta)(\xi|\b_{/i}\comp\g_{i}|\bar\eta)
       (\xi|\a_{/i}\comp\b_{i}|\bar\eta) \nonumber \\ -\,
(\xi|\g_{/i}\comp\b_{i}|\bar\eta)(\xi|\b_{/i}\comp\a_{i}|\bar\eta)
       (\xi|\a_{/i}\comp\g_{i}|\bar\eta), \nonumber
\end{eqnarray}
or using graphical notation for $(\xi|\a|\bar\eta)$
\begin{figure}
\moveright 0.3in \vbox{\epsffile{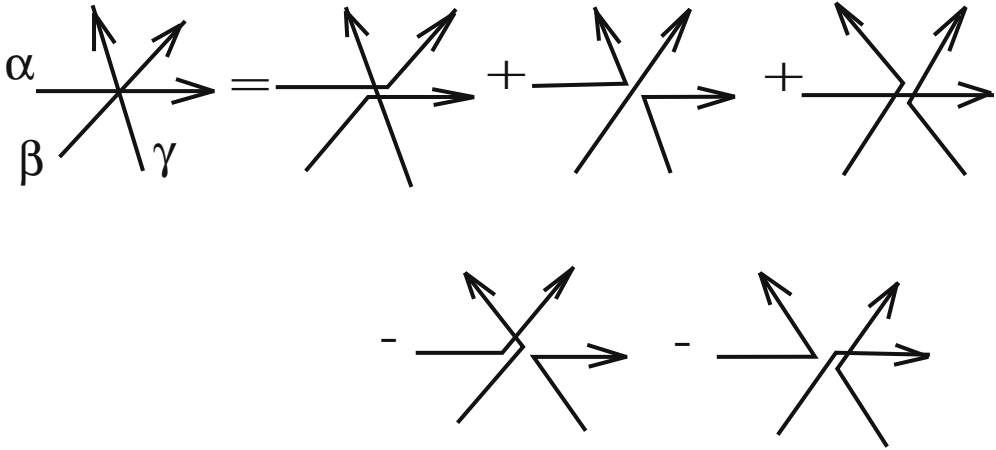}}
\caption{The Mandelstam identity for open curves}
\end{figure}
\item The retracing identity
\[
(\xi|\g_{/s}\comp\a\comp\a^{-1}\comp\g_{s}|\bar\eta)=(\xi|\g|\bar\eta).
\]
\end{enumerate}

So we have introduced the gauge invariant quantities $(\g)$ and
$(\xi|\g|\bar\eta)$. The above identities imply that these quantities are
associated not with loops and curves themselves but with
classes of equivalence of loops and curves. In the case of pure gravity these
classes are called hoops and it seems reasonable to keep this name and
for equivalence classes of curves. Two loops (or curves) belong to the same
equivalence class (or hoop) if they define the same loop quantities for all
fields $\A, \xi, \etabar$.
These hoop variables form an Abelian algebra under the Poisson
brackets and will play a role of ``coordinates'' in the loop representation.

\subsection{Momentum loop variables}

Let us first construct the quantities which do not involve any fermionic
degrees of freedom. We associate
such momentum variables with piecewise strips, i.e. piecewise
ribbons with ends glued \cite{RoSm}.
Inserting the momentum field $\tilde{\cal E}^{a}$ at points of loops,
one can construct the following gauge invariant loop quantities linear
in the field $\E$
\begin{equation}
(\g)^{a}(s) :=
     \Tr\bigl \{ U[\g_{/s}]\,\tilde{\cal E}^{a}(s)\,U[\g_{s}]\bigr \}.
\end{equation}
These quantities are almost what one needs as the momentum variables. As we
have stated, they are gauge invariant but, because of their vector character,
they transform under the action of diffeomorphisms somewhat complicately.
We shall construct the other quantities which are associated
with piecewise strips and which transform under diffeomorphisms as geometrical
objects (i.e. the transformed quantity is also a strip quantity which is
associated to another strip -- a transformed one). In this paper we consider
only the basic, linear in momentum field strip variables which we will again
denote by a letter in paranthesis
\begin{equation}
(S):= \int_{S}d\,s^{ab}(p)\,\eta_{\,abc}(\,\g(p)\,)^{c}(p).
\end{equation}
Here $\g(p)$ is a loop which goes through a point $p$ on $S$, and
$\eta_{\,abc}$ denotes the Levi-Civita tensor density on $\Sigma$. The loop
family $\{\g(p)\}$ is supposed to cover all the strip surface (the loops
$\g(p)$ and $\g(p')$ for different $p, p'$ may coincide). The quantity
defined is the gauge invariant functional on the phase space associated with a
strip $S$
\begin{figure}
\moveright 2.2in \vbox{\epsffile{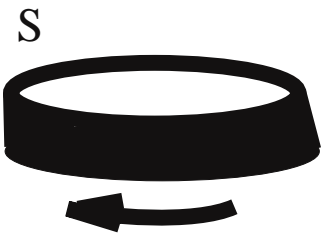}}
\caption{Linear in $\E$ momentum variables are associated with strips.}
\end{figure}

The gauge-invariant loop quantities involving fermionic momentum fields are
\begin{eqnarray}
(\xi|\g|\tilde\pi) := \Tr\bigl \{ \xi\,U[\g]\,\tilde\pi \bigr \}
                         \label{mloopvar},\\
(\tilde\omega|\g|\bar\eta) :=
   \Tr\bigl \{ \tilde\omega\,U[\g]\,\bar\eta \bigr \},
                        \eqnum{\ref{mloopvar}a}\\
(\tilde\omega|\g|\tilde\pi) :=
   \Tr\bigl \{ \tilde\omega\,U[\g]\,\tilde\pi \bigr \}.
                        \eqnum{\ref{mloopvar}b}
\end{eqnarray}
Again $\g$ is a curve with ends marked by the corresponding fermionic
variables. The introduced quantities are represented respectively by
\begin{figure}
\moveright 1.1in \vbox{\epsffile{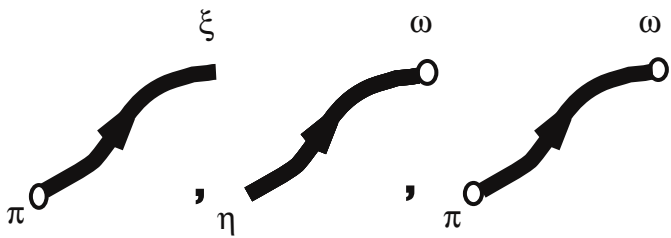}}
\caption{Fermionic momentum variables.}
\end{figure}

\subsection{Loop variables algebra}

The loop variables introduced are functionals on the phase space and
the Poisson algebra they generate can be computed. It is induced by
the Poisson structure on the space of gauge and fermionic fields
((\ref{brackets}) and (\ref{FermBrackets}) respectively). The remaining
part of these section is aimed to
describe the resulting algebra of loop variables in a graphical form.

The brackets of loop quantities with the momentum loop variables
can be obtained by using the following useful expression for the
matrix $U$
\[
U[\g]_{A}^{\;B} =
\int\,ds^{a}
\,U[\g_{/s}]_{A}^{\;C}{\cal A}_{a}(s)_{C}^{\;D}\,U[\g_{s}]_{D}^{\;B}.
\]
So one gets
\begin{equation}
\{\,(S),\,(\b)\,\} = (\g^{(S)}\comp\b) \label{brbegin},
\end{equation}
where $\g^{(S)}$ is the loop from the loop family covering $S$ which
intersect with the loop $\b$. It is tempting to represent this result as
\begin{figure}
\moveright 0.5in \vbox{\epsffile{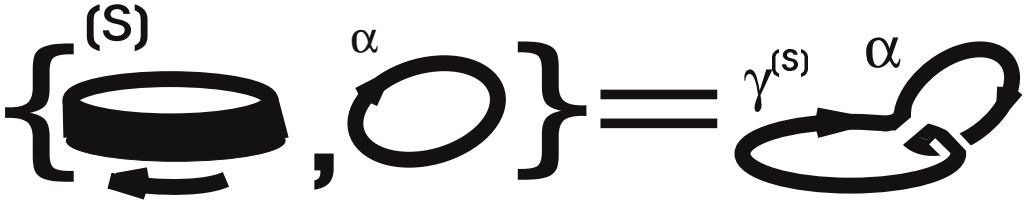}}
\caption{The Poisson brackets between a strip and a loop variables.}
\end{figure}

The brackets of the fermionic momentum variables (\ref{mloopvar}) with the
coordinate loop quantities are given by
\begin{equation}
\int\,d^{3}x\,\{ \,(\xi|\b(x)|\tilde\pi(x)), \,(\xi|\g|\bar\eta) \,\} =
                     (\xi|\b\comp\g|\bar\eta); \label{mloopbr}
\end{equation}
in the right side of this expression $\b$ is the curve from the family
$\b(x)$ whose initial point coincide with the final point of $\g$. There is
the graphical representation for this expression
\begin{figure}
\moveright 0.5in \vbox{\epsffile{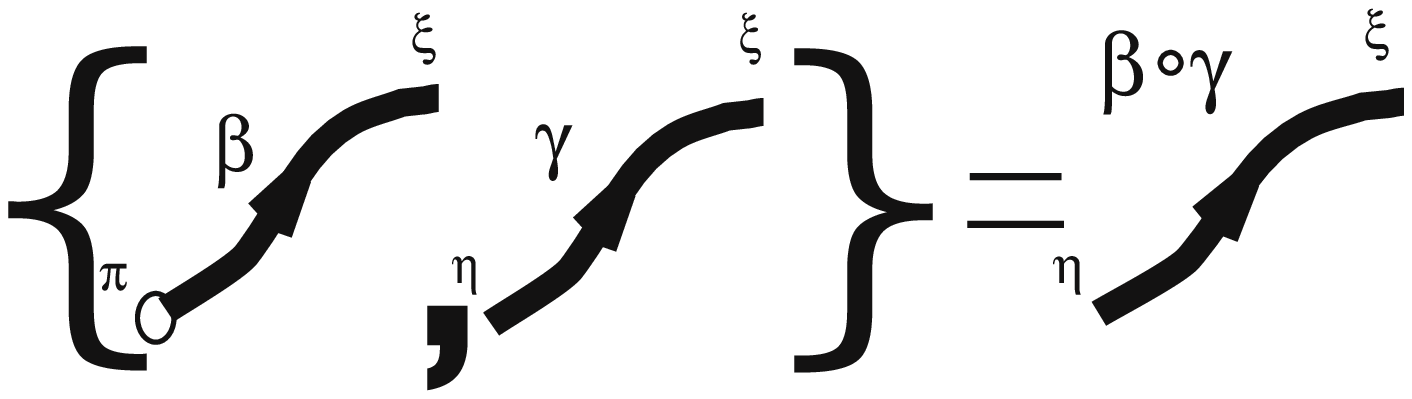}}
\caption{}
\end{figure}
Computing the other brackets one gets
\begin{equation}
\int\,d^{3}x\,\{\, (\tilde\omega(x)|\b(x)|\bar\eta),\,(\xi|\g|\bar\eta)\,\} =
              (\xi|\g\comp\b|\bar\eta);  \eqnum{\ref{mloopbr}a}
\end{equation}
\begin{figure}
\moveright 0.5in \vbox{\epsffile{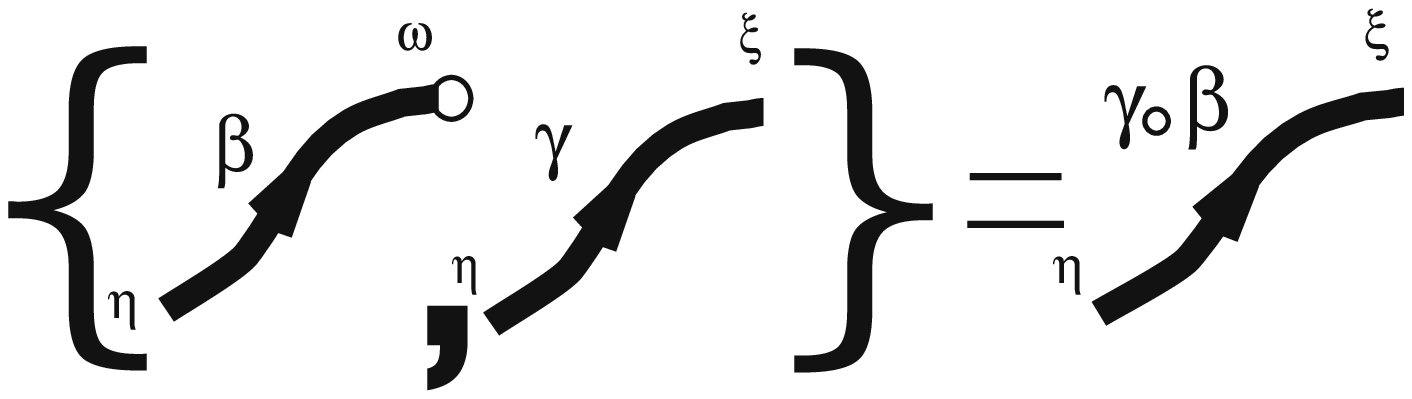}}
\caption{}
\end{figure}
\begin{equation}
\int\,d^{3}x\,d^{3}y\,\{ \,(\tilde\omega(x)|\b(x;y)|\tilde\pi(y)),\,
    (\xi|\g|\bar\eta)\,\} =
    (\xi|\g\comp\b|\pitil)\,+\,(\omegatil|\b\comp\g|\etabar);
    \eqnum{\ref{mloopbr}b}
\end{equation}
\begin{figure}
\moveright 0.1in \vbox{\epsffile{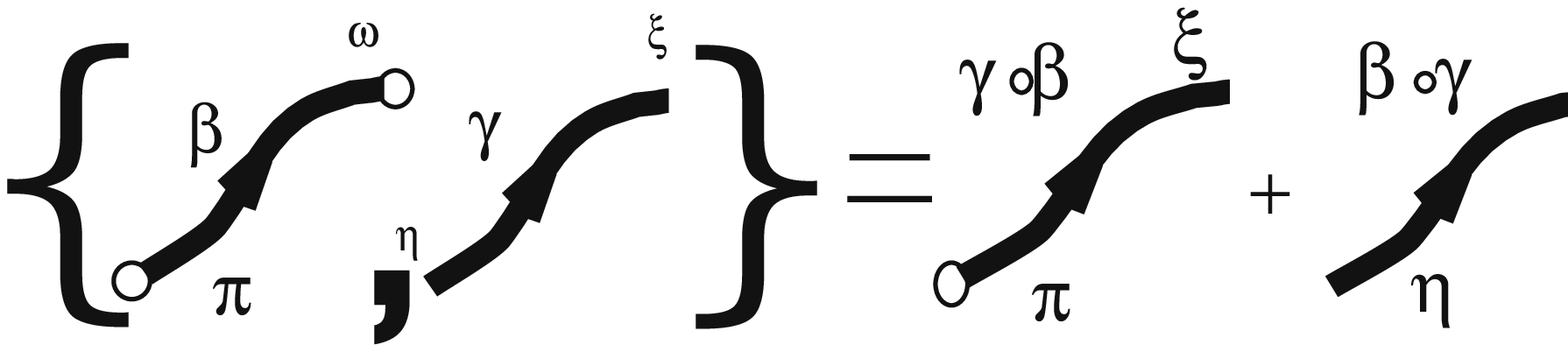}}
\caption{}
\end{figure}

So, as we have seen, the algebra of introduced ``loop'' variables can
be described solely in terms of geometrical objects: loops, curves and strips.
Because of the natural action of the diffeomorphism transformations on the
introduced variables (namely as on geometrical objects), the elements of
the quotient algebra of these variables with respect to the diffeomorphisms'
action have a clear geometrical meaning. They are represented by classes of
diffeomorphism equivalent curves, loops and strips. The algebraic
(induced Poisson) structure on this quotient algebra is given by {\it the
same relations} (\ref{brbegin})-(\ref{mloopbr}) but understanding as
the relations for equivalence classes. This fact helps one to solve the
problem of finding the solutions of the diffeomorphism
constraint in the loop representation.

\section{The Loop Representation}
\label{sec:LoopR}

Constructing the quantum representation for our system we will mostly
follow the programm of Asthekar group (for recent developments see
\cite{AshtRecent}); however, the approach described below is more
"physical" because much importance is attached to the visualization
of all relations.

The program of quantization of generally covariant field theories proposed
in the number of publications (see \cite{AshtRecent} and references therein)
uses the idea to realize the quantum representation space as the
representation space of a configurational variables
Abelian $C^{*}$-algebra. The construction is an infinite-dimensional
generalization of the standart coordinate representation in quantum mechanics:
the space of coordinate representation is the represenation space of the
Abelian algebra of $\hat x$ operators, i.e. $Span\{\ket{x}\}$ where
$\hat{x} \ket{x} = x \ket{x}$. The canonical realization of this space
is the space $L^2(\real, dx)$ of functions $\varphi(x)$ over the spectrum
of $\hat{x}$ operators, i.e. over $\real$. Momentum variables are represented
by derivative operators on $L^2$. The infinite-dimensional case repeats
all these points: the representation space is the space of all continuous
functionals on the spectrum of the configuration variables algebra and
``momentum'' variables naively can be represented by variational derivative
operators.

In order to take advantage of the
standard representaion theory of $C^{*}$-algebras we have to define a
$C^{*}$-algebra of configurational variables.

\subsection{$C^{*}$-algebra of configurational variables}

The most natural candidate for this algebra is the algebra of our
configurational loop variables over complex numbers. Its elements would be
complex even Grassmann numbers (i.e. it would be formed by powers of our
configurational loop variables $(\g)$, $(\xi|\g|\etabar)$) and we would
define an involution as the $\dagger$-operator. One, however, runs into
some problems on this way. First, because of the complexity of the Ashtekar
connection, the parallel transport matrix $U[\g]$ is {\it not} unitary
(it belongs to a larger group
$SL(2,C)\times U(1)$); therefore, our loop variables behave somewhat
complicately under the complex conjugation operation; for example, given any
loop $\g$ there may not
exists such a loop $\g'$ that $(\g)^{*} = (\g')$; . But it is still not the
worst. Because of the non-unitarity of $U(\g)$ the elements $(\g)$ are
{\it not} bounded so the natural sup-norm $\parallel \cdot \parallel$
\begin{equation}
\parallel (\g)\parallel := \sup_{\A}\,|(\g)|  \label{supnorm}
\end{equation}
does {\it not} exists on this algebra. Owing to these facts the case of
complex $A_{a}$ (or the case of Lorentzian general relativity) remained a
problem by the last time. The situation get somewhat changed after the
coherent state transform had appeared \cite{CohTransf}.

Recall that the complex Ashtekar connection $A_{a}$ satisfies
the following {\it reality conditions}\footnote%
 {Although in the given form the reality conditions are non-polynomial in
 $\sigma$, there is a form in which they are polynomial.}
\[
A_{a}+A_{a}^{\dagger} = 2\;\Gamma(\sigma).
\]
Thus, being the complex $SL(2,C)$ connection field, $A$ bears some additional
``unphysical information'' and one can expect that only its ``real'', $SU(2)$
part must play a role in the quantization procedure. This leads to the idea
to consider quantum states $\Psi[A]$ which depend only on the connection
$A_{a}$ and do not depend on its complex conjugate ${\overline A}_{a}$. Such
functionals $\Psi$ are called holomorphic, so the first step to eliminate the
superfluous degrees of freedom from the quantization procedure is to consider
a representation of Lorentzian general relativity in the
space of holomorphic functionals of the Ashekar connection.
Then, as it has been shown in \cite{CohTransf}, there exists an isomorphism
(given by the coherent state transform) between
this space and the space of functionals of  the $SU(2)$ connection.
Therefore, having built a quantum representation of the $SU(2)$ variables
algebra,
one will have the representation of Lorentzian genaral relativity given by the
coherent state transform.

So, according to this scheme,
we have got to construct a representation of the $SU(2)$ variables algebra.
For the case of pure gravity this actually has been done in
\cite{AshtRecent} and our aim is to show that the construction allows a
natural enlargening to the case when the gauge field and the fermionic matter
present.

The $C^{*}$-algebra of the $SU(2)$ configurational variables is described
as follows. It is formed by the same ``loop'' quantities
$(\g), (\xi|\g|\etabar)$ (with the multiplication, additive and $\dagger$-
operations from the Grassmann algebra). The only difference is that the
parallel transport matrix $U[\g]$ becomes now unitary, so it has a bounded
trace and there exists the sup-norm (\ref{supnorm}) for the loop
algebra elements $(\g)$. The unitarity  of $U[\g]$ leads also to the
following simple properties of our algebra generators with respect to
the $\dagger$-operation:
\begin{itemize}
\item Operation $\dagger$ acts on the loop quantities $(\g)$ as simple complex
conjugation and
\[
(\g)^{*}=(\g^{-1}).
\]
\item The action of $\dagger$-operation on the fermionic ``loop''
variables is the consequence of our momentum fields definition
(\ref{momdefine})
and is given by
\begin{eqnarray}
(\sigma(x))(\sigma(y))\,(\xi(x)|\g|\etabar(y))^{\dagger} =
(\omegatil(y)|\g^{-1}|\pitil(x)), \nonumber \\
(\sigma(x))\,(\xi(x)|\g|\pitil(y))^{\dagger} =
(\sigma(y))\,(\xi(y)|\g^{-1}|\pitil(x)), \label{RConditions}\\
(\sigma(y))\,(\omegatil(x)|\g|\etabar(y))^{\dagger} =
(\sigma(x))\,(\omegatil(y)|\g^{-1}|\etabar(x)). \nonumber
\end{eqnarray}
These relations are, in fact, the reality conditions which one should impose
on the fermionic phase space in order to single out its real part.
It is worthwile to note that we have chosen the form in which they are
non-polynomial in the $\E$ variable (because of the presence of $(\sigma)$).
\end{itemize}

Next, let us introduce a norm which makes the algebra of ``loop'' elements a
$C^{*}$-algebra. The algebra is spanned by powers of the ``loop'' generators;
i.e. it is spanned by a set of elements which are
labeled by pairs of points with curves connecting them
(this correspond to a product of $(\xi|\g|\etabar)$ generators) and by
loops (this correspond to a product of different $(\g)$'s). We suppose that
algebra elements depend no more than on a countable set of points
$p_{1},\ldots,p_{k},\ldots,p'_{1},\ldots,p'_{k},\ldots$
and loops
$\g_{1},\ldots,\g_{n},\ldots$,
so arbitrary algebra element $X$ can be written as a functional of the
type\footnote%
 {Here k of this ``loops'' $\g$ are those connecting points $p,p'$ where
 fields $\xi,\etabar$ are taken.}
\[
X = \Phi(\xi_{p_{1}},\ldots,\xi_{p_{k}},\ldots,
 \etabar_{p'_{1}},\ldots,\etabar_{p'_{k}},\ldots,
  U[\g_{1}],\ldots,U[\g_{n}],\ldots).
\]
We call such functionals the cylindical functionals.
Function $\Phi$ is assumed to be a continuous function of all its arguments.

We shall introduce a norm on our algebra by means of the scalar product
which we define in the Appendix. This scalar product has the form (see
(\ref{ScalProd}))
\[
\IP{\Phi_{1}}{\Phi_{2}} = \int\;[\Phi_{1}(\xi,\etabar)]^{\dagger}
\Phi_{2}(\xi,\etabar) \exp{
         \bigl \{ \sum(\xi)^{\dagger}(\xi)+
          \sum(\etabar)^{\dagger}(\etabar)\bigr \}  }\;
          d(\xi,\xi^{\dagger},\etabar,\etabar^{\dagger}),
\]
where the integration is carried over the fermionic fields; the result
does not depend on $\xi,\eta$: it is a
cylindrical functional of connection field (i.e. it depends on $\A$ as a
continuous function $f(U[\g_{1}],\ldots,U[\g_{n}])$). Therefore,
the following function of our algebra elements exists
\begin{equation}
\parallel X\parallel := \sup_{\A\in u(2)}\;\sqrt{\IP{X}{X}},
                     \label{norm}
\end{equation}
which we will show has all the properties of a norm.
\begin{enumerate}
\item First,
\FL
\[
\parallel X+Y \parallel  = \sup\sqrt{\IP{X+Y}{X+Y}} \leq
\sup(\sqrt{\IP{X}{X}} + \sqrt{\IP{Y}{Y}})
\]
\FR
\[
= \parallel X \parallel + \parallel Y \parallel.
\]
\item For two elements $X,Y$ which do not contain fermionic fields the
property
$\parallel XY\parallel\leq\parallel X\parallel\,\parallel Y \parallel$
is obvious. When for example $X$ is a purely fermionic while $Y$ does not
contain fermionic fields the property still holds, because then $\IP{X}{X}$
does not depend on the connection field, as it follows from our scalar product
definition. So we have to check this property only for the case when both $X$
and $Y$ contain fermionic fields; we may always consider one of these
elements, for instance $X$, to be a generator, i.e. a product of the fermionic
``loop'' quantities. In the case when $X = Y$ it can be explicitly checked
that $\parallel X^{2}\parallel = \parallel X \parallel^{2}$.
For instance,
\[
\parallel (\xi|\g|\etabar)^{2}\parallel = \sqrt{4} = \sqrt{2}\sqrt{2} =
             \parallel (\xi|\g|\etabar)\parallel^{2},
\]
so we have the property required. In the opposite case, when $Y$ does
not comprise any curves from those which
compose the generator $X$, the scalar product definition gives
$\IP{XY}{XY}=\IP{X}{X}\IP{Y}{Y}$ and the property is satisfied. So we have
only check it for the case when $Y=X+Y_{1}$, where $Y_{1}$ does not contain
any curves from those composing $X$ (what is equivalent to $\IP{X}{Y_{1}}=0$).
In this case it turns out also that $\IP{X^{2}}{XY_{1}}=0$, so
$\IP{(X+Y_{1})X}{(X+Y_{1})X}=\IP{X^{2}}{X^{2}}+\IP{X}{X}\IP{Y_{1}}{Y_{1}}$,
and, because any fermionic generator has the property that
$\IP{X^{2}}{X^{2}}=(\IP{X}{X})^{2}$, we finally have that
$\IP{(X+Y_{1})X}{(X+Y_{1})X}=\IP{X}{X}(\IP{X}{X}+\IP{Y_{1}}{Y_{1}})=
\IP{X}{X}\IP{Y}{Y}$.
\item Next, we have to check the property $\parallel XX^{\dagger} \parallel =
\parallel X \parallel^{2}$ which makes an algebra with involution a
$C^{*}$-algebra.
Because for the loop quantities this property is obvious,
it should be checked only for the fermionic generators.
The explicit calculation gives
\FL
\[
\parallel (\xi|\g|\etabar)(\xi|\g|\etabar)^{\dagger}\parallel =
\sqrt{\IP{(\xi|\g|\etabar)(\xi|\g|\etabar)^{\dagger}}
         {(\xi|\g|\etabar)(\xi|\g|\etabar)^{\dagger}}} = \sqrt{4}
\]
\FR
\[
        = \parallel (\xi|\g|\etabar)\parallel^{2}.
\]
For the elements of higher orders in the configurational fermionic quantities
the required property follows from the fact that for any two such quantities
$X,Y$ $\IP{XY}{XY}=\IP{X}{X}\IP{Y}{Y}$.
\end{enumerate}

So the algebra of configurational ``loop'' variables becomes an Abelian
*-algebra with norm $\parallel \cdot\parallel$ (which satisfies the relation
$\parallel A\,A^{\dagger}\parallel = \parallel A\parallel^{2}$) and
we can take a completion to obtain a
$C^{*}$-algebra of configurational ``loop'' variables.

\subsection{Representation space}

Having the $C^{*}$-algebra of configurational variables we are at the
point to implement the standard representation theory. According to Gelfand an
Abelian $C^{*}$-algebra is isomorphic to the algebra of all continuous
functions on its spectrum. Let us give
the description of the spectrum of our loop variables algebra. Denote by $\F$
the space of all (satisfying the certain boundary conditions) fields
$\A_{aA}^{\;B}(x), f^{A}(x), g^{A}(x)$ where $\A_{a}\in u(2)$ and $f, g$ are
complex spinor fields ({\it non}-Grassmann-valued) which take values in
fibres $F$ of some $\G$-bundle over $\Sigma$. The corresponding
space quotient
by the gauge transformations will be $\FG$ where $\G=SU(2)\times U(1)=U(2)$.
Then each point of $\FG$ defines a linear homomorphism $\omega$ from the loop
variables algebra to $\Complex$ (a character) as follows:
\[
\omega_{\A,f,g}\bigl ((\g)\bigl ) := \Tr(\exp{\oint_{\g}\A}),
\]
\[
\omega_{\A,f,g}\bigl ((\xi|\g|\etabar)\bigr ) := \Tr(f\;\exp{\int_{\g}\A}\;g).
\]
The spectrum is the set of all characters so we have that the points of $\FG$
distinguish the elements of our algebra
spectrum; it is easy to show that $\FG$ is dence (in Gelfand topology) in the
spectrum, so we will denote the later by $\SP$. This space becomes a quantum
configurational space of our theory. As in the case of pure gravity
its limit points are distributions which we shall regard as generalized
fields (in the sense of Dirac's $\delta$-function). We will denote the
generalized fields by the same symbols $\A, f, g$; so $\SP=\{\A,f,g\}$.

Thus, the space of ``loop'' algebra representation is the space $C^{0}(\SP)$
of all continuous functions over $\SP$.
This space, however, is too large to define integral and differential calculus
on it. The construction of a smoler space, measure and differential calculus
on this smoler space has been proposed by Ashtekar {\it et al}
\cite{AshtRecent}. They proposed to regard the quantum configuration space
of an infinite-dimensional case as {\it the projective limit} of
finite-dimensional configurational spaces of gravity on floating lattices.
Then the representation space becomes the space $Cyl(\SP)$ of cylindrical
functionals over the algebra spectrum. By a cylindrical functional on $\FG$
we understand a map $\Psi$ of the form
\[
\Psi = \Phi(f^{1},\ldots,f^{k},\;g^{1},\ldots,g^{m},\;
{\cal P}\exp{\int_{\g_{1}}\A},\ldots,{\cal P}\exp{\int_{\g_{n}}\A})
\]
\[
\Psi:(F^{k}\times F^{m}\times {\cal G}^{n})\to \Complex.
\]
One can develop a calculus on $Cyl(\SP)$ by means of
the projective limit from the finite-dimensional
configurational spaces \cite{AshtRecent}.
Thus, one can introduce a measure $\mu$ on $Cyl(\SP)$;
this gives rise to the Hilbert representation space $L^{2}(\SP, \mu)$.

In order to construct the {\it loop representation} we
choose a certain basis
in the representation space $Cyl(\SP)$ so that the loop
variables become in this basis simple operators which can be interpreted in
terms of operators of creation and annihilation. The idea is very similar
to one,
which is used to define, for example, the momentum representation in quantum
mechanics. One chooses the basis formed by all proper states $\ket{p}$ of
the momentum operator (the corresponding waive-functions are
$\sim\;\exp{i\,px}$) and defines all the operators by their action on states
from this basis. We introduce the basis of ``loop'' states
which in some sense are proper states of momentum loop operators
and the loop operators become the operators of creation and annihilation of
loops and curves in this basis.

It is convinient to use the Dirac's notations and
denote the following functionals in our space by Dirac's kets
$$\ket{\a} = \Biggl \{
  \lower1em
  \vbox
    {
    \hbox
      {
      $\Tr(f\;{\cal P}\exp{\int_{\a}{\cal A}}\; g) \quad {\rm or}$
      \bigskip
      }
    \hbox
      {
      $\Tr({\cal P} \exp{\oint_{\a}{\cal A}})$
      }
   }  $$
depending on whether $\a$ has ends or not; and in a similar way a
``multiloop'' state is
$$\ket{\a, \b} = \Biggl \{
  \lower1em
  \vbox
    {
    \hbox
      {
      $\Tr(f\; {\cal P} \exp{\int_{\a}{\cal A}} \;g)
       \Tr(f\; {\cal P} \exp{\int_{\b}{\cal A}} \;g)$
      \bigskip
      }
    \hbox
      {\qquad $\cdots$}
   }
   \quad ,{\rm etc.}$$
The order in which loops are taken to compose a multiloop state is not
important.

These states form the basis\footnote%
 {Although, in the absence of a scalar product in the representation space,
 it is note quite legitimate to call the set of states which we will
 introduce a basis,
 we use this term because there {\it exists} a scalar product
 in which the set of states is a basis. Note also, that this basis is not
 countable.}
in the representation space and we
will call them n-loop states. The configurational loop variables become
operators
of multiplication and
the Dirac's notations allow us to express their action simply by
$$(\hat{\g}) \ \ket{\a} = \ket{\a, \g}$$
$$(\xi | \hat{\g} | \eta) \ \ket{\a} = \ket{\a, \g}.$$
In a similar manner they act on states containing more "loops".
We see, therefore, that if one thinks about the state
$\ket{\a_{1}, \a_{2}, \cdots, \a_{n}}$ as about a state containing n "loops",
then the action of the configurational operators consists in simply the
adding of one more "loop" to a state. It is tempting to regard these
operators as ``creation'' operators.
Thus, the basis can be obtained acting
on a cyclic vector (unity functional) by the creation ``loop'' operators.

Let us exemine the basis introduced
more thoroughly. Due to the identities satisfied by Wilson functionals
this basis is overcomplete. Its elements are linear dependent, so some of
them may be rewritten as linear combinations of others. For example,
any three-loop state may be first realized as the state with three
loops intersecting at a point and then reduced to the sum of states
containing two and one loop
\begin{figure}
\moveright 0.2in \vbox{\epsfbox{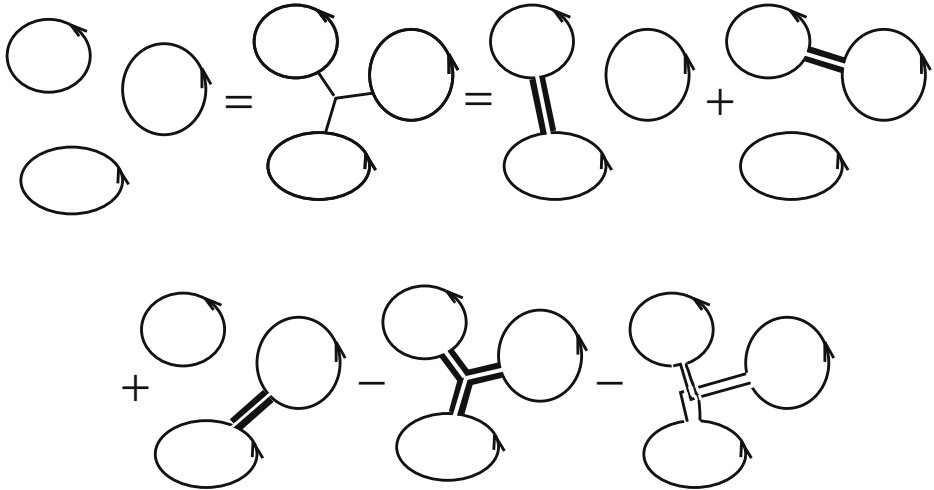}}
\caption{The reduction of a three loop state.}
\end{figure}
Thus, as it has been found by Gambini and Pullin \cite{PuUnified}, any
n-loop state may be reduced to a linear combination of two- and
one-loop states.

Consider then a state containing two loops and one open curve. Repeating
the above procedure, we may reduce this state to a linear combination
of states containing one loop, one open curve and merely open curve states
\begin{figure}
\moveright 0.2in \vbox{\epsfbox{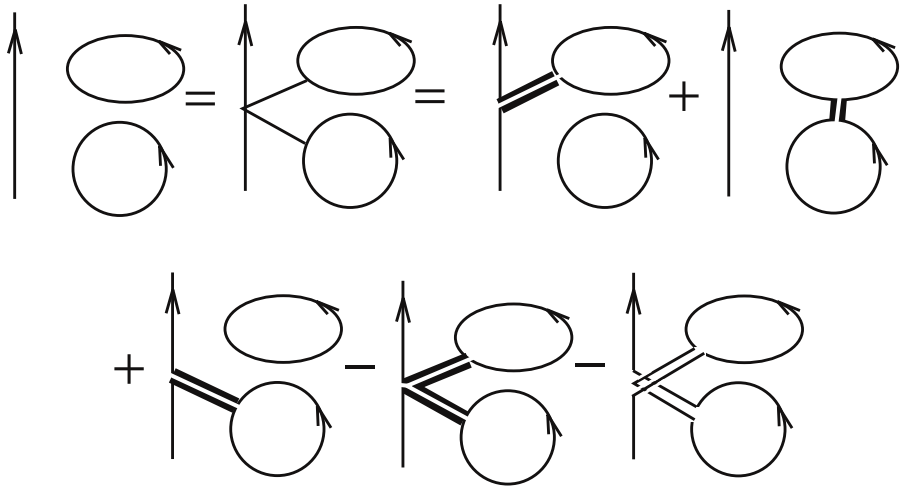}}
\caption{The open curve state reduction.}
\end{figure}

Thus, any state containing n curves and m loops may be reduced to a
linear combination of states containing n curves and one loop or n-curve
states (the number of ends in a state cannot be reduced). The set of
irreduceble elements of our basis consists of states
\begin{itemize}
\item n curves no loops,  n = $0, 1, \cdots$,
\item n curves one loop,  n = $0, 1, \cdots$,
\item two loops.
\end{itemize}
Having this ``loop'' basis in the representation space we are ready to define
the action of other operators by defining it on the
basis elements.

\subsection{Momentum operators}

In the last part of this Section we construct the representation of
the classical momentum ``loop'' variables in the space considered, i.e.
we build operators
\[(\hat S),\quad(\xi|\hat \gamma|\tilde \pi),\quad
(\tilde \omega|\hat \gamma|\bar \eta),\quad
(\tilde \pi|\hat \gamma|\tilde \omega),\]
so that their commutational relations coincide to the first degrees in
$\hbar$ with the Poisson brackets of their classical analogs. Note that we
represent the Poisson brackets by {\it commutational} relations even though
the variables involve Grassmann fields.

The advantage of the built representation is that we have simultaneously
two equivalent descriptions of operators' action. The first, visual one
is based on a graphical representation of states and operators, i.e. on
dealing with Dirac's kets $\ket{\a}$. The second description is based
on representing states as functionals of generalized fields and operators
act in the space of functionals. In this later one there is a naive way to
define the momentum operators; one
shold just use the corresponding classical expressions and replace all
momentum fields by the functional derivative operators.
The resulting operators will act in the space of functionals of generalized
fields $\Psi[\A, f, g]$\footnote%
 {The ``right'' and ``left'' functional derivatives here have nothing to do
 with the Grassmann properties: this is a useful convention which implies the
 rules by which our fermionic momentum operators act on the loop states.
 For example, it is easy to see that
 $(\xi|\hat\g|\lvect{\pitil})\circ(\xi|\a|\etabar)=(\xi|\g\circ\a|\etabar)$ and

%% FOLLOWING LINE CANNOT BE BROKEN BEFORE 80 CHAR
$(\xi|\a|\etabar)\circ(\rvect{\omegatil}|\hat\g|\etabar)=(\xi|\a\circ\g|\etabar)$
 (we omited the integration over the momentum fields in this formula), so the
 right and left vectors over our momentum operators mean simply that the
 $\rvect{\omegatil}$ operator glues a curve to the right side of a curve in
 our ordering (i.e. to the begining point) while $\lvect{\pitil}$ glues a
 curve to the left (to the final point). This also explains the usage of
 $(\cdot|\g|\cdot)$ symbols for our loop quantities: one simply has
 to replace $(\pitil,\xi)$ or $(\etabar,\omegatil)$ pair by the composition
 operation $\circ$.}
\begin{eqnarray}
\E^{a}\to{\hat{\E}}^{a},\quad({\hat{\E}}^{a}\circ\Psi)[\A] :=
           i\;{\delta\over\delta\A_{a}}\Psi[\A];  \nonumber \\
\pitil_{A}\to{\hat{\pitil}}_{A},\quad({\hat{\pitil}}_{A}\circ\Psi)[f,g] :=
           i\;{\lvect{\delta}\over\delta f^{A}}\Psi[f,g];    \\
\label{FuncDer}
%% FOLLOWING LINE CANNOT BE BROKEN BEFORE 80 CHAR
\omegatil^{A}\to{\hat{\omegatil}}^{A},\quad({\hat{\omegatil}}^{A}\circ\Psi)[f,g]
        := i\;{\rvect{\delta}\over\delta g_{A}}\Psi[f,g].  \nonumber
\end{eqnarray}
No problems arise there with operator ordering because functional derivative
operators commute when they act at different space points. These construction
leads to the well defined operators, whose action on the introduced basis
vectors can be described graphically. Moreover, this latter graphical
description can be used as an alternative definition of the ``momentum''
operators. We shall give the result in the both descriptions.

First, let us define the ``strip'' operator, which describes gauge
degrees of freedom. It is represented
by the following operator
\begin{eqnarray}
(\hat S)\circ\ket{\a} := i\,\ket{\g^{S}\circ\a},
\end{eqnarray}
where $\g^{S}$ is the loop from the loop family covering $S$ which intersects
with the loop $\a$ (when there is no intersection of $\a$ with the strip the
result is zero). The graphical representation of this operator is
\begin{figure}
\moveright 0.5in \vbox{\epsfbox{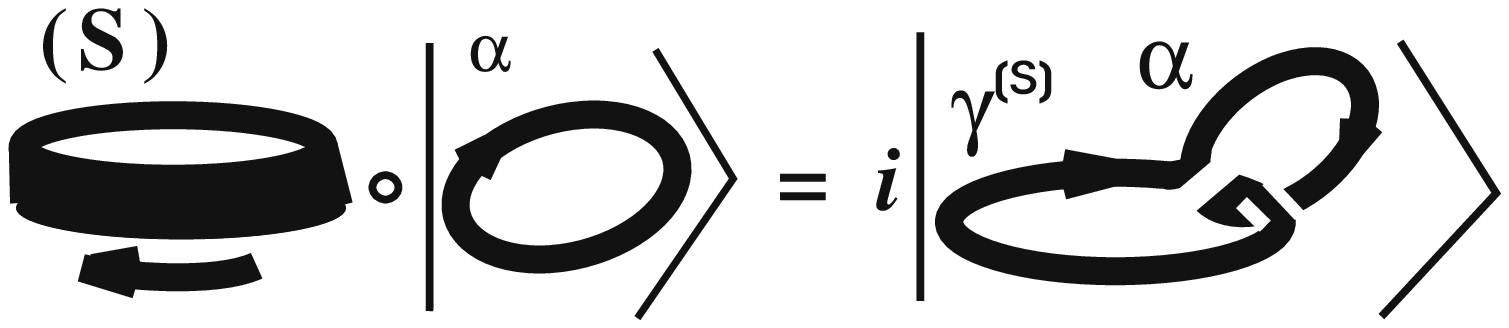}}
\caption{}
\end{figure}
\noindent i.e. the operator adds the loop to a one loop state and glues
these loops in the only way compatible with their orientation. As we have
stated above, the ``loop'' states are the ``proper'' states of our momentum
operators in the sense that the result of their action on a $n$-loop state
is also a $n$-loop state (when the strip intersects more than one loop from
the state the result will be the sum of $n$-loop states).

Let us define the fermionic ``momentum'' operators. The construction is
straightforward from the form of their Poisson brackets with the
configurational variables (one can also obtain the following expressions using
functional derivative operators from (\ref{FuncDer})). We shall define
\begin{equation}
\int\,d^{3}x\,(\xi|\hat \g(x)|\tilde \pi(x))\circ\ket{\a} :=
       i\,\ket{\g \comp \a}.
\end{equation}
Here we integrated over the initial point of the curve $\g$ in order to have
the
same density at the right and the left sides of the expression and the curve
$\g$ at the right side is that one from the family $\g(x)$ whose final point
coincides with the initial point of $\a$. The
graphical description of this operator is
\begin{figure}
\moveright 0.5in \vbox{\epsfbox{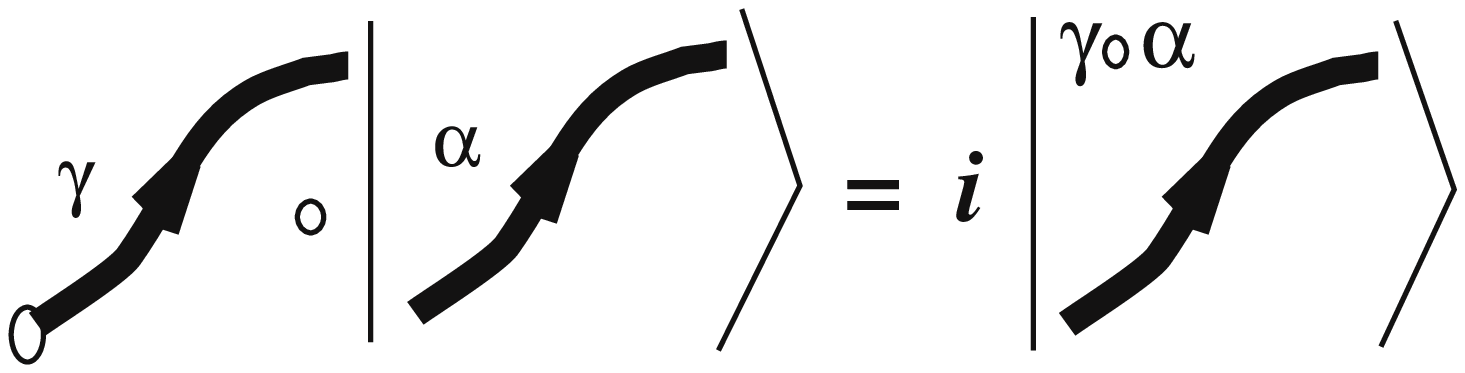}}
\caption{}
\end{figure}
\noindent i.e. it prolonges $\a$ by adding the corresponding curve to the
final point. This, linear in the fermionic momentum field operator does not
change the number of open ends in a state. The other linear in the momentum
field operator is defined in a similar way
\begin{equation}
\int\,d^{3}x\,(\tilde \omega(x)|\hat \g(x)|\bar \eta)\circ\ket{\a} :=
        i\,\ket{\a \comp \g};
\end{equation}
\begin{figure}
\moveright 0.5in \vbox{\epsfbox{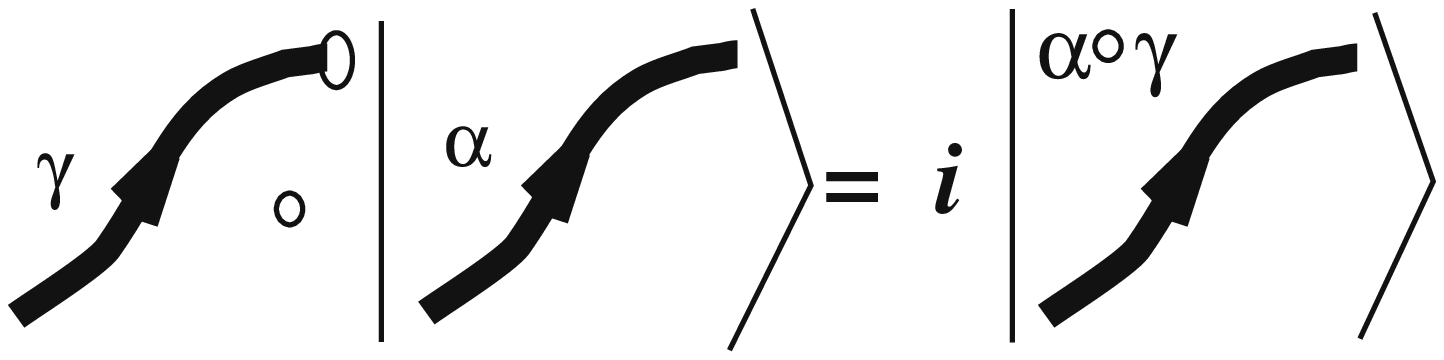}}
\caption{}
\end{figure}
\noindent
the only difference with the previous operator is that this one adds the
corresponding curve to the initial point of $\a$. And, finally, the operator
quadratic in the fermionic momentum fields is defined as
\begin{equation}
\int\,d^{3}x\,d^{3}y\,(\tilde \omega(x)|\hat \g(x;y)|\tilde \pi(y))\circ
\ket{\a}
 := (i)^{2}\;\ket{\g \comp \a};
\end{equation}
\eject
\begin{figure}
\moveright 0.5in \vbox{\epsfbox{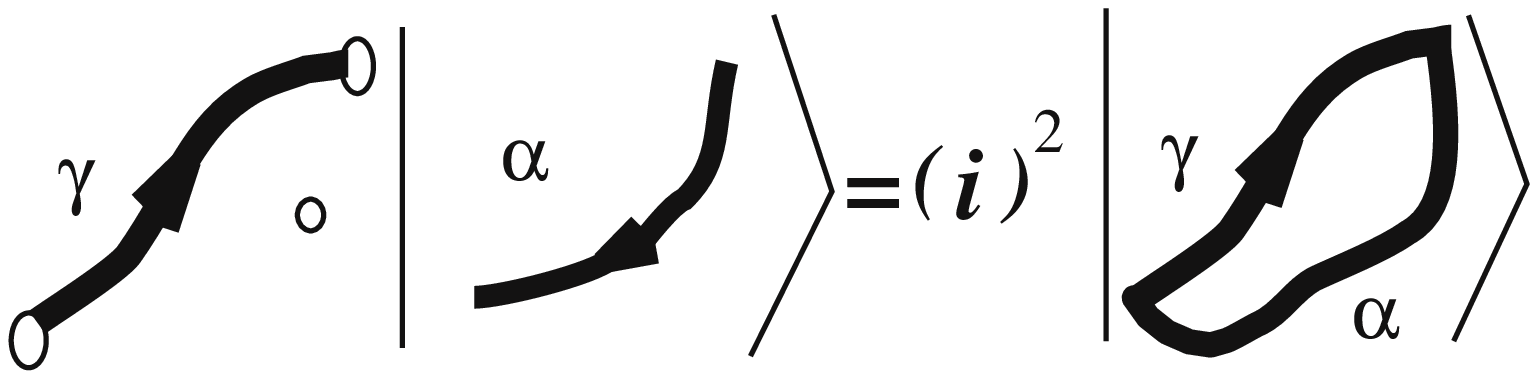}}
\caption{}
\end{figure}
\noindent
That is, the operator glues two open ends of a curve, converting it into a
loop. Acting on more complicated states this operator also reduces (by two)
the number of open ends in a state. The result is the sum of states; in each
of these states the operator glues together two open ends (of different kind),
and the sum goes through all possible pairs of open ends.

Restoring the $\hbar$ factor in all relations one can easy check that the
above defenitions really give a representation of the classical algebra, i.e.
that the commutators among defined operators
(scalled by the factor $i\hbar$) turn into their classical analogs when
$\hbar$ goes to zero.

\bigskip
\bigskip

There are few words left to be said about the solving of the diffeomorphism
constraint in our approach. This is particularly simple while describing the
states and operators graphically, because the elements of quantum
algebra are described in terms of geometrical objects and diffeomorphism
constraint is represented in quantum case by a generator of transformations of
these objects. Commutational relations are also written in terms of
geometrical objects and the relations similar in the form hold for all
representatives of the diffeomorphism equivalent classes of objects.
In order to pick up the physical
states, i.e. to solve the constraint, one should find a representation of the
``physical'' operators which lie in the corresonding quotient algebra. It is
easy to see that in the
approach described this will be the representaion in the space of
equivalence classes of loops and curves and all ``physical'' operators
will act on classes
of diffeomorphic equivalent objects. There is also another, more rigorous
approach for solving the diffeomorphism constraint, which is based on the
description of our quantum operators as operators in the space of functionals;
this approach is given by the averaging procedure from \cite{AshtRecent}.

\section{Discussion}

We have constructed the representation for our quantum system in which
the classical loop variables became operators in the ``loop'' space. The
representation we built differs from the loop representation of pure gravity
in the following important points:
\begin{enumerate}
\item Unlike loops describing pure gravitational exitations, loops and curves
of the unified theory are oriented.
\item The momentum loop operators (corresponding to the gauge as well as to the
fermionic degrees of freedom) act
on the ``loop'' states merely prolonging these ``loops'' in the only compatible
with their orientation way.
\end{enumerate}

One can propose an interesting classification of the constructed ``loop''
operators in terms of creation and annihilation operators.  We have seen that
the operators corresponding to the configurational loop variables act by adding
a ``loop'' to a state, so it is natural to regard them as creation operators.
This terminology is especially good for the operator represented by a curve
because open ends of a curve in our formalism correspond to fermions
\footnote{ Or rather to the fermionic degrees of freedom because of the lack of
interpritation in terms of particles when no background structure presents.}.
It can easily be shown that this operator actually creates a pair of
``fermions''
of different charge sign. Indeed, one can define the charge of a quantum state
as the eighenvalue of the charge operator for this state. Classical charge is
the generator of gauge transformations; it is given by the quantity
$$i\,Q =
     -\int_{\Sigma}d^{3}x
        \,C(x)\,(\xi^{A}(x)\pitil_{A}(x)+\etabar^{A}(x)\omegatil_{A}(x)),$$
where $C(x)$ is an arbitrary (real, integrable) function.
One can define the quantum charge operator with the regularization procedure
of point splitting. The result is a well defined operator
which acts only on the open ends of curves in a state. Each final point on
a curve gives $-1$ while initial points give $+1$. Thus, the result of this
operator's action on {\it any} state in our representation is
zero\footnote{This is what
one would expect from the demand of gauge invariance.}. This means
that our ``fermions'' are born only in pairs with their ``anti-particles''and
that all the states in our representation are electrically neutral.

The operators corresponding to the quantities linear in momentum fields do
not change neither number of loops nor number of open ends in a state; in this
sense the ``loop'' states are the ``eighenstates'' of these operators. And
finaly,
there are momentum operators of higher orders which change the number of
``loops'' in a ``loop'' state; they create or annihilate
``loops'' depending on a state they act on. This completes the description
of quantum kinematics for our unified system.

The important problem which has not been discussed in this paper is the
construction of a scalar product in the representation space. There are
several points in our scheme where a scalar product is nesessary.
First, having defined the ``loop'' operators with respect to the ``basis'' of
``loop'' states, one should have a scalar product
in which the set of ``loop'' states becomes a basis. Second, in order to
have a Hilbert representation space we need a scalar product with respect to
wich the space $Cyl{\SP}$ is complete. And, finally, one needs a scalar
product so that the properties of our operators with respect to the Hermitian
conjugation coincide with the corresponding properties of the classical
quantities with respect to the $\dagger$-operation. This collection of
problems for the system considered is discussed in \cite{HolomRepr}.

Let us conclude by pointing out a possible physical meaning of the
formalism obtained. It describes the unified theory, i.e.
the gravitational and electromagnetic fields enter the formalism
only in a certain combination. On the quantum level exitations of these
fields are described by loops and curves so the formalism predicts that there
do not exist pure gravitational or pure electromagnetic quantum exitations
and these fields appear always only {\it together}.

\section{Acknowledgments}

I am very grateful to Yuri Shtanov
for helpful discussion, comments, and criticism. I thank Carlo Rovelli and
Thomas Thiemann whose critics was very useful. I would especially
like to thank professor P.I.Fomin whose encouragement made this work possible.

\appendix
\section*{}

Let us introduce a scalar product in the space of cylindrical functionals of
fermionic fields. We call a functional $\Phi(\xi,\etabar)$ on the Grassmann
algebra cylindrical if it depends on
fields $\xi,\etabar$ taken no more than in a countable set of points
\[
\Phi = \Phi(\xi_{p_{1}},\ldots,\xi_{p_{k}},\ldots,
            \etabar_{p'_{1}},\ldots,\etabar_{p'_{k}},\ldots);
\]
here we denote $\xi_{p}=\xi(p)$. Then, having the involution $\dagger$,
we can define a
scalar product of two cylindrical functionals $\Phi_{1},\Phi_{2}$ as
\FL
\[
\IP{\Phi_{1}}{\Phi_{2}} :=
\int\;[\Phi_{1}(\xi,\etabar)]^{\dagger}
\Phi_{2}(\xi,\etabar)\;\exp{\sum_{p}(\xi_{p}^{\dagger})_{A}(\xi_{p})^{A} +
                \sum_{p'}(\etabar_{p'}^{\dagger})^{A}(\etabar_{p'})^{A}}
\]
\FR
\begin{equation}
\prod_{p}d\xi_{p}^{\dagger}\wedge d\xi_{p}\wedge
d\xi_{p}^{\dagger}\wedge d\xi_{p}
\prod_{p'}d\etabar_{p'}^{\dagger}\wedge d\etabar_{p'}
\wedge d\etabar_{p'}^{\dagger}\wedge d\etabar_{p'}; \label{ScalProd}
\end{equation}
\[
\IP{1}{1} := 1.
\]
Here $\Phi^{\dagger}$ is to be understood as a cylindrical functional of the
fields $\xi^{\dagger},\etabar^{\dagger}$. The two sums and products are taken
over the points $p,p'$ which the functionals $\Phi_{1},\Phi_{2}$ depend on. The
integral here is the integral over fermionic Grassmann variables; it acquires
a sense if we define the following basic operations on the Grassmann algebra
\begin{eqnarray}
\int\;\xi^{A}\xi^{B}d\xi\wedge d\xi := \epsilon^{AB}; \nonumber \\
\int\;(\xi^{\dagger})_{A}(\xi^{\dagger})_{B}d\xi^{\dagger}\wedge
               d\xi^{\dagger} := \epsilon_{AB}; \nonumber \\
\int\;\etabar_{A}\etabar_{B}d\etabar\wedge d\etabar := \epsilon_{AB};
                                                \nonumber \\
\int\;(\etabar^{\dagger})^{A}(\etabar^{\dagger})^{B}d\etabar^{\dagger}
             \wedge d\etabar^{\dagger} := \epsilon^{AB}, \nonumber
\end{eqnarray}
and
\begin{eqnarray}
\int\;d\xi\wedge d\xi := 0;\quad
\int\;d\xi^{\dagger}\wedge d\xi^{\dagger} := 0; \nonumber \\
\int\;d\etabar\wedge d\etabar := 0;\quad
\int\;d\etabar^{\dagger}\wedge d\etabar^{\dagger} := 0. \nonumber
\end{eqnarray}
Thus, the result of integration of any functional $\Phi$ is defined as
proportional to the
coefficient near the highest order non-zero term in the expansion of $\Phi$ by
powers of the fermionic fields. Then the integration in (\ref{ScalProd}) gives
a
complex number, so this expression really defines a scalar product in the
space of cylindrical functionals.

Finally, let us illustrate the definition on the following simple examples.
The explicit calculation yields
$$\IP{(\xi|\g|\etabar)}{(\xi|\g'|\etabar)} = \Biggl \{
     \lower1em
      \vbox{\hbox{$0,\quad \g\not =\g'$}
            \hbox{}
            \hbox{$2,\quad \g=\g',$}}
$$
\begin{equation}
\IP{(\xi|\g|\etabar)(\xi|\g|\etabar)^{\dagger}}
{(\xi|\g|\etabar)(\xi|\g|\etabar)^{\dagger}} = 4. \nonumber
\end{equation}
\eject

\end{document}